\documentclass[12pt,leqno]{article}
\usepackage{amsmath,amsfonts}
\usepackage{graphicx}
\usepackage{amsthm,amssymb,bm}
\usepackage{setspace}

\newtheorem {defi}{Definition}
\newtheorem {teo} {Theorem}
\newtheorem {prop}[teo] {Proposition}
\newtheorem{rmk}{Remark}

\newcommand{\bomega}{\bm{\omega}}

\begin{document}
\title{Superintegrable 3-body systems on the line}
\author{Claudia Chanu \, Luca Degiovanni \, Giovanni Rastelli \\ \\ Dipartimento di Matematica, Universit\`a di Torino. \\ Torino, via Carlo Alberto 10, Italia.\\ \\ e-mail: claudiamaria.chanu@unito.it \\ luca.degiovanni@gmail.com \\ giorast.giorast@alice.it }
\maketitle

\bigskip

\begin{abstract}
We consider classical three-body interactions on a Euclidean line depending on the reciprocal distance of the particles and admitting four functionally independent quadratic in the momenta first integrals. These systems are multiseparable, superintegrable and equivalent (up to rescalings) to a one-particle system in the three-dimensional Euclidean space. Common features of the dynamics are discussed. We show how to determine quantum symmetry operators associated with the first integrals considered here but do not analyze the corresponding quantum dynamics. The conformal multiseparability is discussed and examples of conformal first integrals are given. The systems considered here in generality  include the Calogero, Wolfes, and other three-body interactions widely studied in mathematical physics. 
\end{abstract}

\paragraph*{I. INTRODUCTION \\}

The analysis of dynamical systems of points on one-dimensional manifolds is a classical subject which has assumed a particular relevance in the last years, in mathematics as well as in physics, with a special attention to the integrable cases. An overview can be found in \cite{OPI,OPII}. Prominent among all is the Calogero-Marchioro-Moser system \cite{Cal,CalMar,Mos}, that is also maximally superintegrable, i.e. admitting $2n-1$ functionally independent global first integrals \cite{R-Woj}, where $n$ denotes the degrees of freedom, which provides matter of study since its discovery. Less known is the Wolfes system \cite{Wol}, where the potential is the superposition of the Calogero system and of a genuine three-body interaction. The Wolfes system is known to be integrable and separable. It has been generalized in \cite{Que}. Due to the fact that it is equivalent to the Calogero system (see Section V), it is also maximally superintegrable. Superintegrability and maximal superintegrability are objects of great interest in modern mathematics and physics, not only because additional first integrals can allow the determination of the trajectories of the systems, but also because differential and algebraic relations between the first integrals themselves enlighten deep features of integrability of differential systems in general. In many cases, superintegrable systems are obtained from multiseparable ones, i.e. systems whose Hamilton-Jacobi equation admits separation of variables in several distinct coordinate systems.  Indeed, for systems with $n$ degrees of freedom the separability in more than one coordinate system often implies the existence of more than $n$ functionally independent quadratic first integrals. For example, in \cite{BlaSer} the Benenti systems, a subclass of the separable systems, are systematically employed for generating maximally superintegrable systems in any finite dimension. The relations between the systems of above and those considered here have not yet been investigated.  Another possible application of multiseparability consists in setting boundary conditions for multiseparable quantum systems by using the coordinate surfaces of different coordinate systems for the same problem.  From the classical multiseparable systems it is always possible, at least in Euclidean spaces and for natural Hamiltonians, to obtain multiseparable quantum systems. Both the Calogero and Wolfes potentials are particular cases of the general one considered in the present paper: a three-body potential on the line with interactions depending on the reciprocal distance of the points. In general, natural Hamiltonian $n$-body systems on the line can be represented as  one-point Hamiltonian systems in the $n$-dimensional Euclidean space. In the present paper we adopt this point of view (differently from \cite{Koz,Per}), and we restrict our analysis to a three-body potential depending on the reciprocal distances only. Moreover, we require  that the potential has the form obtained in \cite{Calnoi,Josh1,SmiWin,SmiWinerr}. A potential in this form was shown in the cited papers to be separable in five distinct coordinate systems and to admit four functionally independent first integrals, quadratic in the momenta. By writing the potential as a function of the distances between the points on the line, we find the  potential obtained in \cite{Koz,Per} and we prove that the Wolfes system is not only completely integrable but also superintegrable and multiseparable. Furthermore, it appears to be related to a larger class of similar systems. In Section II the definitions of superintegrability and multiseparability are given and basic properties of separable systems are recalled. In Section III we briefly describe a fundamental superintegrable and multiseparable potential in $\mathbb{E}_3$. In Section IV we derive the three-body interactions on a line described by a potential equivalent to the previous one and, consequently, multiseparable and superintegrable. In Section V several well known potentials are obtained as particular cases of this fundamental potential and some new are explicited. In Section VI conformal multiseparability is discussed for the systems under consideration. In Section VII we consider very shortly the quantum system corresponding to the classical general one studied in the previous Sections. Some of the results presented here are known in the literature; we provide here a unified approach to the matter with particular emphasis on multiseparability and superintegrability.

\paragraph*{II. SUPERINTEGRABILITY AND MULTISEPARABILITY \\}

We adopt the following definition of superintegrability (see also \cite{KalMil2,R-Woj}),

\begin{defi}
A $n$-degrees of freedom Hamiltonian system is {\bf superintegrable} if it is Liouville integrable and admits more than $n$ functionally independent first integrals. Moreover, the system is called {\bf maximally superintegrable} if it admits $2n-1$ independent first integrals and {\bf quasi maximally superintegrable} if there are $2n-2$ functionally independent first integrals of it. \end{defi}

We require that the Hamiltonian and the first integrals of the above definition are {\it globally} defined, i.e.  they are defined everywhere on the configuration manifold with the exception of a subset of singular points which can be assumed closed and of zero-measure. Since the Hamiltonians and the first integrals considered in this paper are polynomial in the momenta, the simple definition of globality of above satisfies all our needs. 
For example, the Kepler Hamiltonian and its first integrals are for us globally defined in $\mathbb{E}_3$.

For $n=3$ quasi maximally superintegrable systems are also called minimally super-integrable \cite{Eva}.

We call {\bf separable} those Hamiltonians and Hamiltonian systems whose Hamilton-Jacobi equation is integrable by additive separation of variables. As in the classical St\"ackel theory of separaton of variables, we consider here point-transformations of coordinates only. If the Hamiltonian $H$ is natural (kinetic energy plus a scalar potential), then a $n$-dimensional system is separable in orthogonal coordinates if and only if it is a St\"ackel system \cite{Sta, Sta1}. St\"ackel systems of natural Hamiltonians are geometrically characterized by symmetric Killing two-tensors associated with first integrals quadratic in the momenta and in involution according to the following theorem \cite{KalMil1,Ben1,Eis}:

\begin{teo}
The separability of a $n$-dimensional natural Hamiltonian system with potential $V$ in orthogonal coordinates is equivalent to the existence on the configuration Riemannian manifold, of a $n$-dimensional space $\mathcal{K}$ of symmetric Killing two-tensors $\mathbf{K}$, pairwise commuting with respect to Schouten brackets, with common eigenvectors and such that $d(\mathbf{K}\cdot dV)=0$. The orthogonal coordinates are determined by the eigenvectors of the $\mathbf{K}$.
\end{teo}

We recall that the Schouten brackets $[,]$ of two contravariant tensors $A$ and $B$ of order $a,b$ respectively is the symmetric tensor of order $a+b-1$ whose components are defined by
$$
[A,B]^{ij\ldots km}=\frac 1a A^{(li\ldots j}\partial_lB^{k\ldots m)}-\frac 1b B^{(li\ldots j}\partial_lA^{k\ldots m)}
$$
where $(,)$ denote the symmetrization of the indices. The separable coordinate hypersurfaces are orthogonal to the eigenvectors of $\mathbf K$ (the existence of these surfaces is equivalent to the normality of the eigenvectors). The set of the coordinate hypersurfaces is called an {\bf orthogonal separable web}. Any parametrization of it locally defines orthogonal separable coordinates. Separable webs can be grouped in families, according to their geometrical features. In the Euclidean space $\mathbb{E}_3$, for example, orthogonal separable webs are all made of confocal quadrics. We consider equivalent two webs made of the same kind of confocal quadrics, in this way $\mathbb{E}_3$ admits $11$ distinct separable orthogonal webs \cite{Eis}.
\begin{defi}
The Hamiltonian $H$ is {\bf multiseparable} if it is separable in at least two distinct webs.
\end{defi}

The separability of a $n$-dimensional system in several distinct orthogonal separable webs implies the existence of  $n+1\leq r\leq 2n-1$ linearly independent Killing tensors. Each of these Killing tensors $\mathbf K=K^{ij}\partial_i\odot \partial_j$ (where $\odot$ denotes the symmetrized tensorial product) generates a local quadratic first integral $$H_K=\frac{1}{2}K^{ij}p_ip_j+V_K,$$ where $dV_K=\mathbf K\cdot dV$ ($\mathbf K$ is considered as a a linear operator on one-forms). If more than $n$ of them are functionally independent and globally defined, the system is superintegrable. Most of the known superintegrable systems with first integrals that are polynomials of second degree in the momenta,  are obtained as multiseparable systems (they are often called {\bf quadratically superintegrable}). 

\paragraph*{III. A NOTEWORTHY ``SUPER'' POTENTIAL\\}

Let $(x,y,z)$ be Cartesian coordinates in $\mathbb{E}_3$. Let us consider the Hamiltonian
\begin{equation}\label{zero}
H=\frac{1}{2}(p_x^2+p_y^2+p_z^2)+V
\end{equation}
where
\begin {equation}\label{one}
V=\frac{F(y/x)}{x^2+y^2}.
\end {equation}
It is known \cite{Calnoi,SmiWin,Eva} that $H$ is separable in all rotational separable webs around the $z$ axis, namely circular cylindical, spherical, parabolic, spheroidal prolate and spheroidal oblate \cite{MooSpe}. This property follows from
\begin{prop} \label{teo2}
In any rotational orthogonal coordinate system $(q^1,q^2,q^3)$ with rotational axis $z$ and angle of rotation $\psi=q^3$, the potential (\ref{one}) takes the form of a St\"ackel multiplier
\begin{equation}\label{eq2}
V=g^{33}F(q^3),
\end{equation}
where $g^{ii}$ are the components of the metric tensor in $(q^1,q^2,q^3)$.
\end{prop}
\begin{proof}
Every rotational coordinate system $(q^1,q^2,q^3)$ around the $z$ axis can be transformed to coordinates $(x,y,z)$ by the change of variables
$$
\left\{
\begin{array}{ll}
x&=f_1(q^1,q^2)\cos q^3\\
y&=f_1(q^1,q^2)\sin q^3\\
z&=f_2(q^1,q^2).
\end{array}
\right.
$$
We immediately have $g^{33}=1/{f_1}^2$, and $V=F(\tan q^3)/{f_1}^2$. Since $F$ is a generic function, $V$ can be written as (\ref{eq2}).
\end{proof}

The coordinate systems in which the Hamiltonian (\ref{zero}) is separable are generated by a $5$-dimensional linear space of Killing tensors. Correspondingly, five quadratic first integrals can be constructed and result globally defined \cite{Calnoi, SmiWin}. However, only four of them are functionally independent (in \cite{Calnoi, SmiWin} it is erroneously reported that the five quadratic first integrals are all functionally independent \cite{SmiWinerr}). The system with Hamiltonian (\ref{zero}) is hence quasi maximally superintegrable; it is unknown if the system is maximally superintegrable for all $F$. For some particular forms of $F$ a fifth functionally independent constant of motion can be constructed, thus making the system maximally superintegrable (see, for example, Remark 6 and \cite{Josh1,Win}).

By using cylindrical coordinates $(r,\psi,z)$, with rotational axis $z$, and by indicating with $(p_r,p_\psi,p_z)$ their conjugate momenta, the potential  (\ref{one}) becomes
\begin{equation}\label{iv}
V=\frac{F(\psi)}{r^2}
\end{equation}  
and the five quadratic first integrals take the form \cite{Calnoi,SmiWin}
\begin{eqnarray*}
H=H_0&=& \frac{1}{2} \left(p_r^2+\frac 1{r^2}p^2_\psi+p^2_z\right)+\frac {F(\psi)}{r^2}, \\
H_1&=&\frac{1}{2}p_\psi^2+F(\psi), \\
H_2&=&\frac{1}{2}p_z^2,\\
H_3&=&\frac{1}{2}\left[(rp_z-zp_r)^2+\left(1+\frac{z^2}{r^2}\right)p_{\psi}^2\right] +\left(1+\frac{z^2}{r^2}\right)F(\psi),\\
H_4&=&\frac{1}{2} \left(zp_r^2 +\frac{z}{r^2}p_{\psi}^2-rp_rp_z\right) + \frac{z}{r^2}F(\psi).
\end{eqnarray*}
It is easy to check their polynomial dependence \cite{SmiWinerr}
\begin{equation}\label{dipendenza}
H_0(H_3-H_1)-H_2H_3-H_4^2=0.
\end{equation}
The first three integrals allow the separation of the system in cylindrical coordinates. In particular, the conservation of $H_2$ implies that $p_z$ is constant. By setting $H_i=h_i$ we obtain the three differential equations:
\begin{eqnarray}
\dot{z}^2&=&p^2_z=2h_2 \nonumber\\
\dot{r}^2&=&p_r^2=2(h_0-h_2)-\frac{2h_1}{r^2},\label{eqsep}\\
\dot{\psi}^2&=&\frac{1}{r^4}p_\psi^2=2\frac{h_1-F(\psi)}{r^4}. \nonumber
\end{eqnarray}
Necessarily $h_2\geq 0$, and the motion of the system occurs only for $(r,\psi)$ such that 
$$ 
(h_0-h_2)r^2\geq h_1 \geq F(\psi)
$$ 
The additional first integrals $H_3$ and $H_4$ provide a relation between $r$ and $z$:
$$
h_2r^2=(h_0-h_2)z^2-2h_4z+h_3-h_1
$$
that allows to find the radial law of motion without integrating (\ref{eqsep}), for initial conditions such that $h_2\neq0$.

When $h_2=0$, the motion becomes planar. Then, from (\ref{eqsep}) we obtain
\begin{equation}\label{er1}
\dot{r}^2=2h_0-\frac{2h_1}{r^2}
\end{equation}
that determines the radial component of the motion. It is remarkable that the radial motion is the same for all potentials (\ref{one}) and that $\dot{r}=0$ for $r^2=h_1/h_0$ only, where a simple zero occurs. Therefore, no closed trajectory is possible except for the particular case  $h_0=h_1=h_2=0$ when the trajectory is on a circle.

\paragraph*{IV. THREE-BODY SYSTEMS ON THE LINE\\}

We consider a natural Hamiltonian system of three points on a line, with positions $x^i$ and momenta $p_i$. Without loss of generality we can assume that the points have the same mass. This system is equivalent to the natural Hamiltonian system in $\mathbb{E}_3$ with coordinates $(x^i)$ of the point $\mathbf{x}=(x^1,x^2,x^3)$. We require that the interaction depends only on the distance between the points, including in particular three-body interactions. The potential has, consequently, the form
\begin{equation} \label{Vdist}
V=U(X_i), \qquad  i=1,2,3
\end{equation}
where
$$
X_i=x^{i}-x^{i+1}, \quad i=1,2,3 \quad \mbox{(mod $3$)}.
$$
Therefore, the system is invariant under the translation $\bomega=(1,1,1)$, as a consequence of the invariance of the momentum of the center of mass.
The form (\ref{iv})
$$
V=\frac{F(\psi)}{r^2}
$$
of $V$ can be referred to any rotational coordinate system where $\psi$ represents the angle of the rotation and $r$ is the distance from the axis of rotation. In the following, the axis of rotation is parallel to the vector $\bomega$ and the transformation between $(x^i)$ and the cylindrical coordinates $(r,\psi,z)$ is therefore given by
\begin{equation}
\left\{
\begin{array}{ll}
r\cos \psi &=\frac1{\sqrt{2}}(x^1-x^2)\\
r\sin \psi &=\frac1{\sqrt{6}}(x^1+x^2-2x^3)\\
z&=\frac 1{\sqrt{3}}(x^1+x^2+x^3).
\end{array}\label{Coo}
\right.
\end{equation}
We want to determine  the function $U$ in (\ref{Vdist}) such that the potential is of the form  (\ref{iv}). %or, equivalently,  which $F$ in (\ref{iv}) represents a three-body interaction on a line depending on the  reciprocal distance of the particles only (\ref{Vdist}). 
\begin{teo}\label{teogen}
The most general function $V$ of the form (\ref{iv}) can be written as
\begin{equation} \label{gen}
V=\sum _i\frac 1{X_i^2}F_i\left(\frac {X_{i+1}}{X_i},\frac{X_{i+2}}{X_i}\right) \quad i=1, \ldots, 3 \mbox{ (mod }3),
\end{equation}
where $F_i$ are generic functions of two variables.
\end{teo}
\begin{proof}
Let $\bomega $ be a unit vector parallel to the axis of rotation, and let $$\mathbf {r}=\mathbf{x}-\frac{(\mathbf{x}\cdot\bomega)\bomega}{\omega^2}$$ be the radial vector from the axis of rotation to the point $\mathbf{x}$. The potential $V$ is manifestly invariant with respect to $\bomega$, i.e. it does not depend on $z$. Therefore, the requirement that $V$ in (\ref{Vdist}) has the form (\ref{iv}) means that $r^2 U$ depends only on the angle, i.e. it is invariant with respect to $\mathbf{r}$. Since
$
\mathbf{r}(r^2)=2r^2, 
$
we have
$$
\mathbf{r}(r^2 U)=2r^2 U+ r^2 \mathbf{r}(U)
$$
and the invariance condition is 
$$
\mathbf{r}(U)=-2U.
$$
In the variables $X_k$ the previous equation is equivalent to
$$
X_i\frac{\partial}{\partial X_i}U(X_j)=-2U(X_j)
$$
whose solution is of the  form 
$$
U=\frac 1{X_1^2}F_1(\frac {X_{2}}{X_1},\frac{X_{3}}{X_1})
$$
that without loss of generality can be written more symmetrically as (\ref{gen}).
\end{proof}

\begin{rmk}
Due to the relation $X_1+X_2+X_3=0$ the function (\ref{gen}) can be rewritten in the two equivalent forms 
$$
V =\frac{1}{X_1^2}F\left(\frac{X_2}{X_1}\right), \qquad
V = \frac{F(X_2/X_1)}{X_1^2+X_2^2}.
$$
\end{rmk}

\begin{rmk}
Potentials of form (\ref{gen}) are known to be integrable by quadratures only for zero values of total energy and momentum since the paper by Kozlov and Kolesnikov \cite{Koz, Per}.
\end{rmk}
Therefore,

\begin{teo}
The three-body system on the line with Hamiltonian
$$
H=\frac{1}{2} (p_1^2+p_2^2+p_3^2)+\sum _i\frac 1{X_i^2}F_i\left(\frac {X_{i+1}}{X_i},\frac{X_{i+2}}{X_i}\right), \quad i=1, \ldots, 3 \;(\mathrm{mod}\;3),
$$
is quasi maximally superintegrable and separable in five types of rotational coordinate systems.
\end{teo} 
\begin{proof}
It follows from Theorem \ref{teogen} and the properties of potential (\ref{one}).
\end{proof}

\begin{rmk}
The potential is also separable in any other separable rotational coordinate system with the same axis and arbitrary origin \cite{Calnoi}.
\end{rmk}

\begin{rmk}
If the points have positive masses $m_i$, a rescaling $y^i=\sqrt{m_i}x^i$ make the metric Euclidean. Then, by setting $X_i=y^i-y^{i+1}$, $i=1,2,3$ (mod $3$), and $\bomega =(1/\sqrt{m_1},1/\sqrt{m_2},1/\sqrt{m_3})$ the above procedure can be repeated \cite{Calnoi}.
\end{rmk}

\paragraph*{V. EXAMPLES\\}
There are several well-known  examples of potentials representing a three-body interaction on the line  that can be written in the form (\ref{gen}): for instance the Calogero inverse square potential \cite{Cal,Calnoi} 
\begin{eqnarray*}\label{VI}
V_I=\frac{k_1}{(x^1-x^2)^2}+\frac{k_2}{(x^2-x^3)^2}+\frac{k_3}{(x^3-x^1)^2}= 
\sum_{i=1}^3 \frac {k_i}{X_i^2},\qquad k_i\in \mathbb R.
\end{eqnarray*} 

Another interesting example is the Wolfes potential describing a genuine three-body interaction \cite{Wol},
\begin{eqnarray*}\label{VII}
V_{II}&=& \frac{k_1}{(x^1+x^3-2x^2)^2}+\frac{k_2}{(x^2+x^1-2x^3)^2}+\frac{k_3}{(x^3+x^2-2x^1)^2}\\
&=&\sum_i \frac{k_i}{(X_i-X_{i+1})^2}\\
&=&
\sum_i \frac{1}{X_i^2}k_{i+1}\left(\frac{X_{i+1}}{X_i}- \frac{X_{i+2}}{X_i} \right)^{-2}, \quad i=1, \ldots, 3 \mbox{ (mod }3).
\end{eqnarray*}
\begin{rmk}
In the  Calogero and Wolfes potentials usually
considered in literature we have $k_1=k_2=k_3$ and these functions are often considered together in a single potential. In this case, written in cylindrical coordinates $(r,\psi,z)$ the functions $F(\psi)$ corresponding to $V_I$ and $V_{II}$ are respectively $k_I\, \sin^{-2} 3\psi$ and $k_{II}\,\cos^{-2} 3\psi$, for suitable constants $k_I$, $k_{II}$. The two kinds of  systems therefore coincide after a rotation in the three-dimensional space.
\end{rmk}
 
\begin{rmk}
It is easy to show that, at least for $n=0,\ldots, 4$, systems with $F(\psi)=k\sin^{-2} 2n \psi$ ($n\neq 0$) and $F(\psi)=k\sin^{-2} (2n+1) \psi$ admit additional irreducible polynomial first integrals of degree $2n+1$ in $(p_r, p_\psi)$, and the same for $F(\psi)=k\sin^{-2} \psi/2n$ ($n\neq 0$), $F(\psi)=k\sin^{-2} \psi/(2n+1)$ (as seen in the previous remark, the functions $sin$ and $cos$ are interchangeable). The coefficients of the polynomials are fairly regular and this allows to conjecture that the properties of above hold for all integers $n$. For example, for $n\geq 0$ a fifth functionally independent first integral for the potential $V=k\left(r\; \sin (2n+1) \psi\right)^{-2}$ seems to be  
$$
H_5=\sum_{\sigma =0}^{n}\sum_{i=0}^{2\sigma+1} {\frac{A_\sigma ^i }{r^{2n+1-i}}\left(\frac{2k}{\sin ^2 (2n+1)\psi}\right)^{n-\sigma}\frac {d^{2\sigma+1-i}\left(\cos (2n+1)\psi\right)}{d\psi^{2\sigma+1-i}} p_r^ip_\psi^{2\sigma+1-i}}
$$
with
$$
A_\sigma^i=\frac{(-1)^{2n-\sigma}}{(2n+1)^{2\sigma+1-i}} \; \left( \begin{matrix} {2n+1} \cr i \end{matrix}\right) \left( \begin{matrix} [(2n+1-i)/2] \cr [(2\sigma+1 -i)/2]\end{matrix} \right),  
$$
where $\left(^a _b\right) =\frac {a!}{b!(a-b)!}$ denotes the Newton binomial symbol and $[a]$ the greatest integer  $\leq a$. This would provide a sequence of maximally superintegrable systems with non trivial polynomial first integrals of any odd degree. The analysis of these cases is in progress.
\end{rmk}  
 
The potential 
$$
V_{III}=\frac{c_1}{x^2}+\frac{c_2}{y^2}
$$
where $(x,y)$ are Cartesian coordinates in the plane, is separable in Cartesian, Polar, Elliptic-Hyperbolic coordinates and in the corresponding cylindrical systems of the space (\cite{Win} and references therein). The function $V_{III}$ represents a superintegrable interaction on a line. Indeed, by a change of coordinates from $(x,y,z)$ to $(x^1,x^2,x^3)$, where the $z$-axis coincides with the line for the origin and parallel to $(1,1,1)$, the potential can be written as
\begin{eqnarray*}
V_{III}&=&\frac{c_1}{2(x^1-x^2)^2}+ \frac{c_2}{6}\frac 1{(x^1+x^2-2x^3)^2} \\
&=&\frac{c_1}{2X_1^2}+\frac{c_2}{6}\frac 1{(X_2-X_3)^2}\\
&=&\frac{1}{X_1^2}\left[\frac{c_1}{2}+\frac{c_2}6\left(\frac{X_{2}}{X_1}- \frac{X_{3}}{X_1}\right)^{-2}\right]. 
\end{eqnarray*}
It can be considered as a mixture of Calogero and Wolfes interactions.
For the potentials $V_I$ and $V_{III}$, cubic first integrals functionally independent of the quadratic ones are known \cite{Ran,Win,R-Woj}, thus making these systems maximally superintegrable. 
After Theorem 4 it is not difficult to build quasi maximally separable potentials on the line, for example
$$
V_{IV}=\sum_{i=1}^{3}\frac{k_i}{X_i^2+X_{i+1}^2}=\sum_{i=1}^3\frac {k_i}{X_{i+2}^2}\left(\frac{X_i^2}{X_{i+2}^2}+\frac{X_{i+1}^2}{X_{i+2}^2}\right)^{-1} \mbox{ (mod }3).
$$
Moreover, by starting from the expression $(\ref{iv})$ of the potential $V$ it is possible to write it as the potential of an interaction among three bodies on the line. Indeed, from (\ref{Coo}) we obtain
$$
\mbox {tan} \, \psi=\frac 1{\sqrt{3}}\left( 2\frac{X_2}{X_1}+1\right) 
$$ 
and
$$
r^2=\frac 43(X_1^2+X_1X_2+X_2^2).
$$
For example, with reference to Remark 6, by expanding $\sin ^2 9\psi$ in powers of $\mbox{tan}\, \psi$ we have, after some computations, 
$$
\frac 1{r^2\sin ^2 9\psi}=\frac{3(X_1^2+X_1X_2+X_2^2)^8}{(X_2-X_1)^2(2X_1+X_2)^2(2X_2+X_1)^2W_1W_2},
$$
where $W_1=(X_2^3-3X_1X_2-6X_1^2X_2-X_1^3)^2$ and $W_2=(X_2^3+6X_1X_2+3X_1^2X_2-X_1^3)^2$. 

\paragraph*{VI. CONFORMAL MULTISEPARABILITY\\}

A natural Hamiltonian $H=G+V$, where $G$ is the geodesic term, is {\bf conformally separable} for the value $0$ of the energy if the Hamiltonian $H/V$ is separable in the usual sense. The separation is then associated with conformal Killing two-tensors of $G$ and conformal first integrals of $H$ in a way pretty similar to the standard separation. Recall that a conformal first integral of $H$ is a function $K$ such that $\{H,K\}=fH$ for some function $f$. If $f=0$ the separation is the standard one. All integral curves of $H=0$ coincide with the integral curves of $H/V=-1$ up to a time rescaling (Jacobi transformation) \cite{HJE}. A necessary and sufficient condition for the conformal separability of $H=G+V$, where $G$ is the Euclidean geodesic term,  in orthogonal conformally separable coordinates of $\mathbb{E}_3$, is that in these coordinates the potential takes the form of a pseudo-St\"ackel multiplier: $V=g^{ii}\phi_i(q^i)$.  
Indeed, the Hamiltonian (\ref{zero}) is conformally separable  in all orthogonal conformally separable  rotational coordinate systems of the three-dimensional Euclidean space with axis of rotation $z$. In fact, (\ref{one}) coincides in these systems with $g^{33}F(\psi)$. The rotational orthogonal conformally separable coordinate systems of $\mathbb{E}_3$ are: tangent spheres, cardioids, inverse oblate, inverse prolate, bi-cyclide, flat-ring-cyclide, disk-cyclides \cite{MooSpe}. More details about conformal separation of Hamilton-Jacobi and Schr\"odinger equations can be found in \cite{HJE,FER,CMl}. In the case of genuine conformal separation, the set of zero energy (or of the value of the energy for which conformal separation holds) is well defined because, if $V$ is a pseudo-St\"ackel multiplier, then $V+h$ is not a pseudo-St\"ackel multiplier for any $h\neq 0$, otherwise the separation would be the standard one \cite{HJE}. For conformal separation the potential function cannot be chosen up to additive constants.  
For instance, the conformal first integral corresponding to separation in cardioids coordinates is
$$
H_c=z(z^2-r^2)p_z^2+2zr^2p_r^2+r(3z^2-r^2)p_rp_z
$$ 
and  
$$
\{H,H_c\}=2H[4zrp_r+(3z^2-r^2)p_z],
$$
which is zero for integral curves contained in the hypersurface $H=0$.

The four functions $(H_1,H_2,H_4,H_c)$ are functionally dependent on $H=0$, with $2H_4(H_1\, H_2+H_4^2)+H_2^2 H_c=0$, but any three of them together with the Hamiltonian are not. It is possible  that there are  systems with more functionally independent conformal first integrals for one fixed value of the energy than for the others, but this seems not to be the case. For all these conformally separable systems, $H, H_1$ and $H_c$ are the quadratic conformal first integrals associated with conformal separation in cardioids coordinates. We remark that $H_c$ is the same for all scalar potentials of the form (\ref{one}) because it does not depend on $F$. Some of the natural systems with potential (\ref{one}) admit a wider set of separable coordinates. In six-spheres coordinates $(u,v,w)$ (obtained as inversion of the Cartesian coordinates), for example, the potential $V_{III}$ becomes
$$
V_{III}=\Delta^2\left(\frac{c_1}{u^2}+\frac {c_2}{v^2}\right), \qquad \Delta=u^2+v^2+w^2
$$
and, because $g^{uu}=g^{vv}=g^{ww}=\Delta^2$, it is a pseudo-St\"ackel multiplier. Then, the potential is conformally separable in these coordinates \cite{FER}. The same property holds in the coordinate systems obtained by inversion from the circular and elliptic cylindrical coordinates (the other coordinates that allow separation of variables). Even if conformal multiseparability do not enlighten new features of the potentials we are considering, the possibility of separation of variables in these new systems can be useful if one wishes to consider perturbations of the potentials. For example, by adding to them for each of the conformally separable coordinate system a perturbative term in the form of $fH$, where $f$ is a pseudo-St\"ackel multiplier, one obtains different integrable systems whose dynamics coincide with the original one for $H=0$. Vice-versa, if $f$ is any function, the dynamics on $H=0$ could be a multiseparable approximation of the perturbed one for "small values" of $H$.

\paragraph*{VII. QUANTIZATION\\}

For $H,H_1,\dots , H_4$ it is easy to build  corresponding quantum symmetry operators by following quantization rules given for example in \cite{SchI,SchII,DuvVal}. At a classical level, symmetric Killing two-tensors $\mathbf{K}_k$ with components $K_k^{ij}$ are associated to quadratic first integrals of the Hamiltonian $H=\frac 12 G^{ij}p_ip_j+V$ of the form $H_k=\frac 12 K_k^{ij}p_ip_j+V_k$, provided some compatibility conditions involving $V$ are satisfied, for suitable functions $V_k$. Correspondingly, self-adjoint differential operators of order two $\hat H_k$ are defined by $\hat H_k \phi=\frac 12\nabla_i(K^{ij}_k\nabla_j \phi)+V_k\phi$ for any wave function $\phi$. The operator corresponding to the Hamiltonian $H$ is the Schr\"odinger (Laplace-Beltrami) operator $\hat H$. Since in Euclidean spaces the Ricci tensor is null, the Robertson condition holds. Therefore, the differential operators associated to Hamiltonians in involution do commute. It follows that a symmetry operator of $\hat H$ corresponds to each quadratic first integral of $H$. If the first integrals are in involution, then the corresponding differential operators commute and $\hat H$ is multiplicatively separable in the same coordinates as $H$. Thus, in analogy with the classical multiseparability previously shown for $H$, we have
\begin{teo}
The Schr\"odinger operator $\hat H$ associated to $H$ admits four distinct second order symmetry operators and is separable in five distinct webs.
\end{teo}
We do not consider here the problem of quantum superintegrability, more on this topic can be found in \cite{GW, KMK} and references therein.
For higher order first integrals the procedure of quantization is less understood.
For cubic first integrals a quantization rule is given in \cite{DuvVal} as follows $P=P_3^{jkl}p_jp_kp_l+P_1^jp_j\longmapsto \hat P=\frac i2\left( -(\nabla_jP_3^{jkl}\nabla_k\nabla_l+\nabla_j\nabla_kP_3^{jkl}\nabla_l)+P_1^j\nabla_j+\nabla_jP_1^j\right)$. We do not develop here the analysis of the quantum systems corresponding to (\ref{gen}). Particular cases are discussed for example in \cite{Cal,Wol,OPII,Que}.

\paragraph*{VIII. CONCLUSION\\}

Several well known  integrable systems are considered as examples of a more general system representing  three points on a line whose dynamics is
quasi maximally superintegrable and multiseparable, and some new are given. Conformal multiseparability is  discussed for the general system considered. A quantum system corresponding to the classical one is shown to be  multiseparable. The analysis of three-body systems  on the line done here is extensible to $n$-body interactions on the line or on  higher dimensional Euclidean manifolds. Again, quasi maximally superintegrable and multiseparable systems are obtained. These systems will be considered in a forthcoming paper. 

\paragraph*{ACKNOWLEDGMENTS\\}

We wish to thank Pavel Winternitz for useful discussions on the topic of this article. 

This research has been partially supported by the project of the Ministero dell'Universit\'a e Ricerca: PRIN 2006-2008.


\begin{thebibliography}{}

\bibitem{Ben1} S. Benenti, J. Math.Phys. {\bf 38} 6578 (1997)

\bibitem{Calnoi} S. Benenti, C. Chanu and G. Rastelli, J. Math.Phys. {\bf 41} No. 7, 4654 (2000)

\bibitem{SchI} S. Benenti, C. Chanu and G. Rastelli, J. Math.Phys. {\bf 43} No. 11, 5183 (2002)

\bibitem{SchII} S. Benenti, C. Chanu and G. Rastelli, J. Math.Phys. {\bf 43} No. 11, 5223 (2002)

\bibitem{HJE} S. Benenti, C. Chanu and G. Rastelli, J. Math.Phys. {\bf 46} 042901/29 (2005) 

\bibitem{BlaSer} S. Blaszak and A. Sergeev, J. Phys. A {\bf 38} (2005)

\bibitem{Cal} F. Calogero, J. Math.Phys. {\bf 10} 2191 (1969)

\bibitem{CMl} C. Chanu, M. Chanachowicz and  R. G. McLenaghan, J. Math.Phys. {\bf 49} 013511 (2008)

\bibitem{FER} C. Chanu and G. Rastelli, IJGMMP {\bf 3} No. 3, 489 (2006)

\bibitem{DuvVal} C. Duval and G. Valent, J. Math.Phys. {\bf 46}, 053516 (2005)

\bibitem{Eis} L. P. Eisenhart,  Ann. Math. {\bf 35} 284 (1934) 

\bibitem{Eva} N. W. Evans, Phys. Rev. A {\bf 41} 5666 (1990)

\bibitem{GW} S. Gravel  and P. Winternitz, J. Math. Phys. {\bf 43} (12) 5902 (2002)  

\bibitem{Josh1} J. T. Horwood, J. Math. Phys. {\bf 48} 102902 (2007)

\bibitem{KalMil1} E. G. Kalnins and W. Miller Jr., SIAM J. Math. Anal.{\bf 11} 1011 (1980) 

\bibitem{KalMil2} E. G. Kalnins, W.  Miller Jr. and G. Pogosyan, Symmetries and Overdetermined Systems of Partial Differential Equations, IMA volumes in Mathematics and its applications. M.Eastwood and W.Miller editors, Springer {\bf 144} 431 (2008)

\bibitem{KMK} E. G. Kalnins, J. Kress and W. Miller Jr., J. Math. Phys {\bf 47} 043514 (2006) 

\bibitem{Koz} V. V. Kozlov and N. N. Kolesnikov, Vestnik Mosk. Univ. Ser. Mat. Meh. No. 6,  88 (1979)

\bibitem{CalMar} C. Marchioro, J. Math.Phys. {\bf 11} 2193 (1970)



\bibitem{Win} I. Marquette  and P. Winternitz, J. Math.Phys. {\bf 48} 012902 (2007)

\bibitem{MooSpe} P. Moon and D. E. Spencer, Field theory handbook, Springer-Verlag, Berlin (1988)

\bibitem{Mos} J. Moser, Adv. Math. {\bf 16} 197 (1975)

\bibitem{OPI} M. A. Olshanetsky and A. M. Perelomov, Physics Reports (Review Section of Physics Letters) {\bf 71} No. 5  313 (1981)

\bibitem{OPII} M. A. Olshanetsky and A. M. Perelomov, Physics Reports (Review Section of Physics Letters) {\bf 94} No. 6  313 (1983) 

\bibitem{Per} A. M. Perelomov, Integrable systems of classical mechanics and Lie algebras, Birkhauser-Verlag, Berlin (1990)

\bibitem{Que} C. Quesne, Phys. Rev. A (3) {\bf 55} No.5, 3931 (1997)

\bibitem{Ran} M. F. Ra\~nada, J. Math.Phys. {\bf 40} 236 (1999)

\bibitem{SmiWin} R. G. Smirnov and P. Winternitz, J. Math.Phys. {\bf 47} 093505 (2006)

\bibitem{SmiWinerr} R. G. Smirnov and P. Winternitz, J. Math.Phys. {\bf 48} 079902 (2007)

\bibitem{Sta} P. St\"ackel, Math. Ann. {\bf 42} 537 (1893)

\bibitem{Sta1} P. St\"ackel, Ann Mat. Pura Appl. {\bf 26} 55 (1897)

\bibitem{R-Woj} S. Woijechowski, Phys. Lett. A {\bf 95} 279 (1983)

\bibitem{Wol} J. Wolfes, J. Math.Phys. {\bf 15} 1420 (1974)

\end{thebibliography}
\end{document}